\documentclass[%
 aip,
 amsmath,amssymb,
 reprint,%
]{revtex4-1}

\pdfoutput=1
\usepackage{graphicx}% Include figure files
\usepackage{dcolumn}% Align table columns on decimal point
\usepackage{bm}% bold math

\usepackage{xcolor}
\usepackage[normalem]{ulem}

\usepackage[utf8]{inputenc}
\usepackage[T1]{fontenc}
\usepackage{mathptmx}

\begin{document}

\title[]{Raster Thomson scattering in large-scale laser plasmas produced at high repetition rate}
% Force line breaks with \\

\author{M. Kaloyan}
 \email{mkaloyan@g.ucla.edu.}
\affiliation{Department of Physics and Astronomy, University of California Los Angeles, Los Angeles, CA 90095, USA 
}

\author{S. Ghazaryan}
\affiliation{Department of Physics and Astronomy, University of California Los Angeles, Los Angeles, CA 90095, USA 
}

\author{C. G. Constantin}
\affiliation{Department of Physics and Astronomy, University of California Los Angeles, Los Angeles, CA 90095, USA 
}

\author{R. S. Dorst}
\affiliation{Department of Physics and Astronomy, University of California Los Angeles, Los Angeles, CA 90095, USA 
}

\author{P. V. Heuer}
\affiliation{
Laboratory for Laser Energetics, University of Rochester, 250 East River Road, Rochester, NY 14623, USA
}%

\author{J. J. Pilgram}
\affiliation{Department of Physics and Astronomy, University of California Los Angeles, Los Angeles, CA 90095, USA 
}

\author{D. B. Schaeffer}
\affiliation{ 
Department of Astrophysical Sciences, Princeton University, Princeton, NJ 08540, USA
}

\author{C. Niemann}
\affiliation{Department of Physics and Astronomy, University of California Los Angeles, Los Angeles, CA 90095, USA 
}

%\date{\today}% It is always \today, today,
             %  but any date may be explicitly specified

\begin{abstract}
We present optical Thomson scattering measurements  of electron density and temperature in a large-scale ($\sim 2$ cm) exploding laser plasma produced by irradiating a solid target with a high energy (5-10 J) laser pulse at high repetition rate (1 Hz). The Thomson scattering diagnostic matches this high repetition rate. Unlike previous work performed in single shots at much higher energies, the instrument allows point measurements anywhere inside the plasma by automatically translating the scattering volume using motorized stages as the experiment is repeated at 1 Hz. Measured densities around 4$\times 10^{16}$ cm$^{-3}$ and temperatures around 7 eV result in a scattering parameter near unity, depending on the distance from the target.
The measured spectra show the transition from collective scattering close to the target to non-collective scattering at larger distances. Densities obtained by fitting the weakly collective spectra agree to within 10$\%$ with an irradiance calibration performed via Raman scattering in nitrogen. 
\end{abstract}

\maketitle

\section{\label{sec:introduction}Introduction}
Thomson scattering (TS) is a powerful first-principles, non-invasive plasma diagnostic that has been applied extensively to characterize 
dense millimeter-scale laser produced plasmas (LPP) encountered in inertial confinement fusion and high-energy density physics research \cite{glenzer1999, froula12}. 
%Scattering in these hot and dense plasmas is dominated by the collective interactions between electrons and ions. 
%
%
There has recently been increasing interest in understanding the less explored late-time evolution of exploding laser-plasmas. These are used, for example, to study the physics of space and astrophysical phenomena in appropriately scaled laboratory experiments~\cite{niemann2014}, for thin film deposition of materials~\cite{leboeuf1996}, and for laser-ablation propulsion research~\cite{phipps2010}. At these late times, the LPP expands to much larger scales of centimeters~\cite{schaeffer_jap} to meters~\cite{heuer2020}, while its density and temperature drop significantly. In order to apply TS in these plasmas, one has to overcome challenges associated with the small scattering cross section, similar to those encountered when scattering off magnetic confinement fusion plasmas~\cite{peacock1969,murmann1992,johnson1999,leblanc2008} or low temperature plasmas~\cite{deregt1998, vds, kono2000,meiden2008,carbone2014}. These include the need for stray light mitigation using a notch filter, an absolute irradiance calibration of the detection branch, and photon averaging over hundreds of laser shots.

We have developed a high-repetition rate TS diagnostic to provide model-independent volumetric measurements of electron density and temperature in large-scale LPPs produced at high shot rate (1~Hz) anywhere inside the plasma, with high spatial ($<$ 1~mm) and temporal (4~ns) resolution. 
Most previous TS systems for single shot laser-plasma experiments have produced measurements at one single point in space and time \cite{schaeffer2012}. Some experiments have employed imaging spectrometers to obtain 1D-spatially resolved data at one single time \cite{gregori2004,schaeffer2016}, while others
have used streaked spectrometers to obtain plasma parameters at a single spatial point as a function of time \cite{glenzer1999}.
By contrast, the instrument described here uses motorized stages to auto-align and translate the scattering volume between laser shots. 
Spatially resolved data is obtained by automatically scanning the scattering volume across a predefined spatial pattern,
as the experiment is repeated thousands of times.
Automated alignment and translation of the scattering volume are made possible by an optical design that optimizes the trade-offs between alignment sensitivity on the one hand and optical extent and photon count on the other. 
This capability to collect volumetric measurements of plasma density and temperature is important for a new generation of high-repetition-rate high-energy laser facilities \cite{hrr1, hrr4, hrr2, hrr3}.

In section \ref{sec:setup} we present a detailed description of the instrument, its optical design, and the notch filter for stray light rejection. 
%The absolute irradiance calibration via Raman scattering is described in section \ref{sec:calibration}.   % 
First proof-of-concept scattering measurements in a relatively unexplored large-scale ($\sim$2~cm), high-energy (5-10~J) LPP are presented in section \ref{sec:results}. For now these experiments were limited to measurements along the blow-off axis due to the low probe beam energy (50 mJ). Electron densities obtained by fitting the weakly collective spectra agree with an absolute irradiance calibration via Raman scattering (section \ref{sec:calibration}) to within 10$\%$, which is well within the accuracy of the density measurement.

\section{\label{sec:setup}Experimental setup}
\begin{figure}[!ht]
    \centering 
    \includegraphics[width=8.5cm]{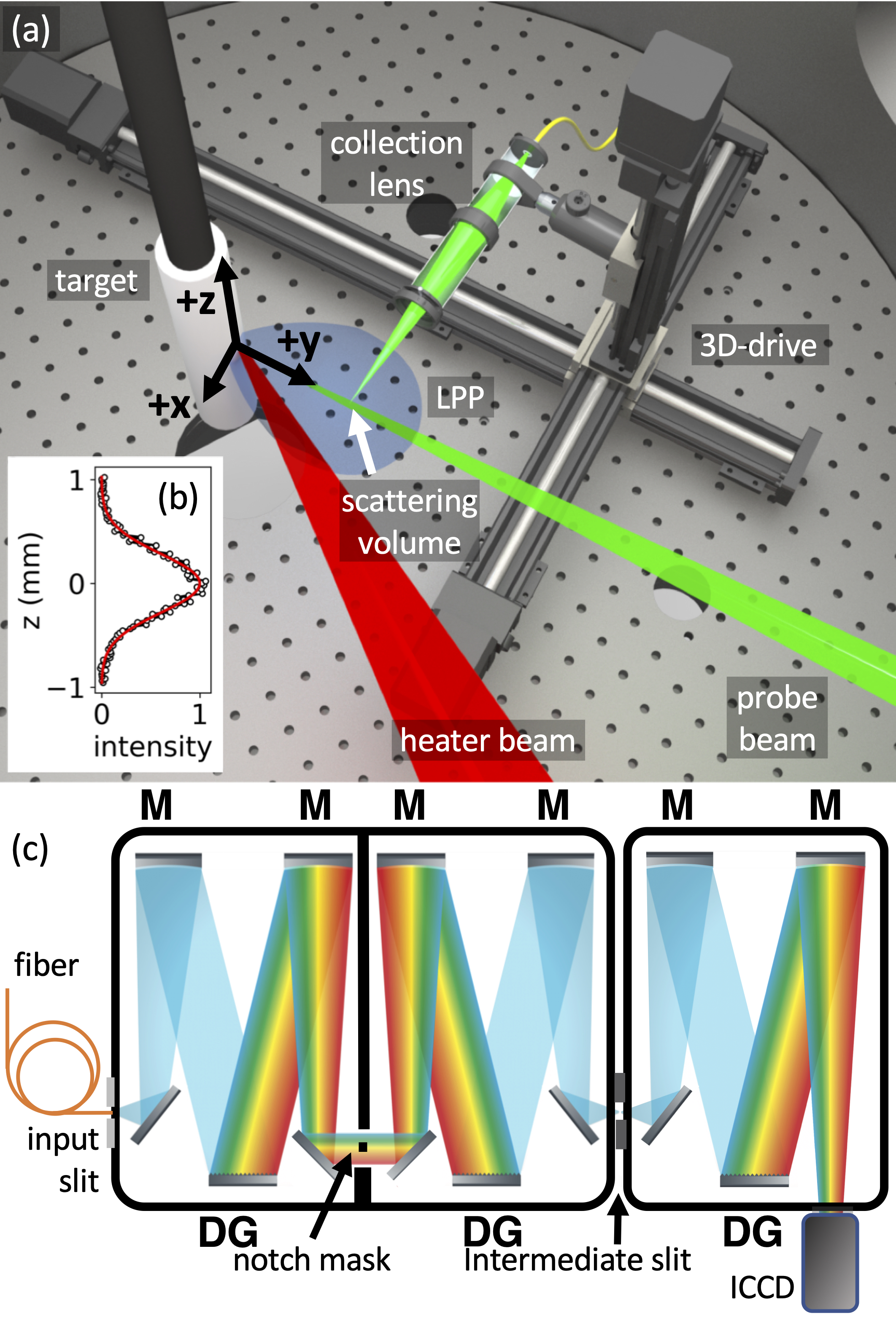}
    \caption{\label{fig:setup} (a) Schematic layout of the experiment showing the target and beam configuration and the collection branch on the 3D-drive. (b) The scattering volume is automatically aligned by scanning the lens and fiber along z while monitoring the Raman scattering intensity off air.  (c) A diagram showing the light path through the triple grating spectrometer used for stray light suppression shows the notch mask between the double-subtractive stages.}
\end{figure}

A schematic of the experimental setup is shown in Fig. \ref{fig:setup}a.
The plasma is created by ablating a solid plastic target with a high energy laser pulse. The laser used to heat the target is a high-repetition rate flashlamp-pumped Nd:glass zig-zag path slab laser  with 
wavefront correction using a stimulated Brillouin scattering phase conjugate mirror \cite{dane}. The laser delivers
an energy up to 10~J in a near diffraction limited beam at 1053 nm with a pulse duration of 14 ns at a repetition rate of 1~Hz.
The output energy is stable to within 5$\%$, and the pulse shape and pointing are stable to within 1$\%$.
The beam is focused onto a cylindrical high density polyethylene (C$_2$H$_4$) plastic target mounted on a 2D stepper motor drive. The target drive is synchronized with the laser, and is rotated and translated in a helical pattern to provide a fresh surface for each shot \cite{schaeffer2018}. %Each target, with a diameter of 25~mm and a length of 30~cm, could be used for more than 10$^4$ laser shots.
The beam is incident on the target at an angle of $34^\circ$ relative to the target normal and is
focused to an intensity around $\sim 10^{12}$ W/cm$^2$ using an f/25 spherical lens with a focal length of 1.0 m.
The LPP explodes normal to the target surface regardless of the angle of incidence. 
A 200 l/s turbo-pump stabilizes the
vacuum pressure during 1 Hz laser ablation at around 7$\times$10$^{-3}$ torr.

A separate pulse from a second diode-pumped Nd:YAG laser is used as the TS probe beam.  This beam delivers an 
energy of $E_{\rm{pulse}} = 50$~mJ in a $\tau_L = 4$ ns pulse at the second harmonic ($\lambda_i=532$ nm), 
and enters the target chamber through an anti-reflection coated flat vacuum window and disposable blast shield exactly along the plasma blow-off axis ($-\hat{y}$).
A slow (f/150) spherical lens with a focal length of 75~cm is used to focus the probe beam to a point 3~cm from the target surface along $\hat{y}$. This configuration produces a cylindrical "pencil" beam of constant 0.4~mm diameter over the 6~cm Rayleigh length in front of the target. 
The probe beam is synchronized to the heater pulse including a variable delay between 50-300~ns, and is 
linearly polarized along the z-axis. 
The probe beam terminates on the target, creating copious amounts of stray light as well as a secondary laser-plasma plume. However, this secondary laser plasma reaches the scattering volume hundreds of ns after the time of the scattering measurement, and consequently does not affect our results. 

Scattered light is collected perpendicular to the probe beam with a 5.0~cm focal length, 12~mm diameter aspheric collection lens into an f/20 collection cone, and is focused at f/5.3 into a single 200~$\mu$m core optical fiber with a numerical aperture of NA~=~0.22. The 40~m long fiber is coupled to the input slit of the spectrometer via a fiber vacuum feed-through.
The scattering volume is defined by the intersection of the probe beam and the projection of the magnified fiber core cross section onto the beam. 
The collection branch is designed to project the fiber core with a 3.8$\times$ magnification onto the probe beam, so that the 0.7~mm diameter
fiber projection exceeds the probe beam size by a factor of $\sim 2$.
At first glance it would seem that the slow f/20 collection cone would result in more than an order of magnitude loss in collection efficiency, compared to a much faster design often used (f/4).
However, it should be considered that the total number of collected photons is proportional to the source optical extent $G' = A'\cdot \Omega'$, where $A'$ is the source area, and $\Omega'=\pi/(2\cdot f\#')^2$ is the collection cone solid angle, where $f\#'=20$ is the collection cone f-number. The larger f-number results in a larger source area because the magnification of the fiber onto the probe beam $|M|= f\#' / f\#$ increases linearly with $f\#'$, while the f-number into the fiber $f\#=5.3$ is fixed.
When scattering off a pencil beam with a diameter smaller than the fiber projection, $G'$ scales inversely proportional to the f-number.
The parameters used here result in $G'=8.5\times 10^{-4}$~rad$\cdot$~mm$^2$, which is still 70$\%$ of the theoretical upper limit set by the optical extent $G$ of the spectrometer f-number and fiber area. 
This acceptable loss in photons significantly reduces alignment sensitivity, 
making it possible to automatically translate the scattering volume along the probe beam using motorized stages.

The collection lens, along with a fused silica blast shield, 3 mm diaphragm, and fiber launch are housed in a 
light-tight 25~mm diameter tube. The assembly is installed on a 3D stepper-motor controlled probe drive inside the vacuum vessel, which can position the scattering volume anywhere inside a 10$\times$10$\times$10$~$cm$^3$ volume in front of the target, with 5~$\mu$m resolution and bidirectional repeatability.
Raman scattering off of air at atmospheric pressure is used to automatically align the collection branch to the probe beam.
Fig. \ref{fig:setup}b shows the measured Raman intensity as a function of vertical lens position ($\hat{z}$).
The data was obtained using a camera exposure of 1~s with the probe laser pulsing at 50~Hz.
The Gaussian fit to the measured profile (red line) has a width of 0.7~mm (FWHM), which corresponds to the convolution of fiber core projection and the 0.4 mm diameter probe beam. The center of the fit marks the optimum lens alignment. 
The vertical beam position is mapped out automatically this way as a function of distance from the target ($\hat{y}$). 
Raman scattering measurements in air confirmed that the collection efficiency remains constant when translating the scattering volume along the probe beam ($\hat{y}$).
Similarly, collection efficiency does not vary when translating the collection lens transversely to the probe ($\hat{x})$ over a range of $\pm$1~cm, as expected given the large mismatch between fiber projection and beam diameter, and the slow f/20 collection cone.

 A triple-grating spectrometer (TGS) with a spatial filter is used to significantly reduce stray light without eliminating the TS signal. 
 %The purpose of the notch is to reduce the stray light as much as possible without cutting  much  of  the  TS.
The spectrometer used is a 0.5~m focal length f/4 Czerny-Turner type imaging spectrometer with three 1200~grooves/mm holographic gratings. Scattered light enters the spectrometer through a 200~$\mu$m  input slit, is collimated and imaged by internal mirrors, and passes through a notch filter and 2.0~mm intermediate slit before reaching the detector. The first two gratings are used in subtractive dispersion and serve as a notch filter to reject stray light. The last grating disperses the light onto the detector. Toroidal mirrors image the input slit onto the notch filter, intermediate slit and detector plane. 
As illustrated in Fig. \ref{fig:setup}c, the notch filter is placed in between the double-subtractive spectrometers. We used an 0.75~mm wide and 50~$\mu$m thick stainless steel mask, which blocks a wavelength range of of 1.5~nm. 
The  total transmission  through the spectrometer without the notch is  measured  to  be  25$\%$ and  is  mostly  determined by the reflectivity of the three  holographic  aluminum-coated  gratings  blazed  at  500~nm. The spectral resolution was measured to be 0.22~nm over a wavelength range of 19.4~nm.

Spectra are recorded using an image intensified charge couple device (ICCD) directly mounted to the output of the third spectrometer stage.
The camera has a {\it generation III} photocathode with a quantum efficiency of 50$\%$ at 532~nm.
The micro-channel plate (MCP) is gated at 4~ns and synchronized to the 4~ns probe pulse with a jitter of less than 100~ps.
We kep the MCP gain fixed at the maximum of of G~=~233, and used 2$\times$2 hardware binning for all images. Even with full gain and binning, the  pixel count is a small fraction of the 16-bit maximum and well within the linear response range. The spectra are further binned over 250 vertical pixels along the fiber and slit and over two pixels horizontally into 256 total bins with 0.076~nm/bin.
The 20~e$^-$ / pixel readout noise with the ICCD cooled to \mbox{-20$^o$C} is negligible with this binning when averaging multiple shots. 

Carbon plasma self emission lines from the LPP are also present in the spectrum that can have intensities higher than the scattering signal.
To eliminate these lines, a plasma-only background spectrum recorded without the probe is subtracted from each scattering spectrum.
An equal number of TS images and plasma-only background images are interleaved by creating the LPP at 1~Hz but pulsing the probe laser on only every other shot (1/2 Hz). 
While the plasma emission and residual stray light background can be subtracted from the data, their shot noise can not.
The total noise is determined by the shot noise of both the TS signal and plasma background. 
Integrating over $n$ laser shots increases the total number of TS counts and background counts equally to $n\cdot N_{TS}$ and $n\cdot N_{bg}$, respectively. 
Assuming Poisson statistics, the noise of each signal component increases with the number of shots as $\sqrt{n}$, as does the signal to noise ratio (SNR)  
\begin{equation}
    \label{snr}
    \rm{SNR}  =  \frac{n \cdot N_{TS}}{\sqrt{n\cdot N_{TS} + n\cdot N_{bg}}}\ \sim\ \sqrt{n}\ . 
\end{equation}
In order to increase SNR we averaged between 100-400 shots (in addition to an equal number of background shots) for each spectrum, depending on the density and the amount of plasma emission.

\begin{figure}[!ht]
    \centering 
    \includegraphics[width=3.4in]{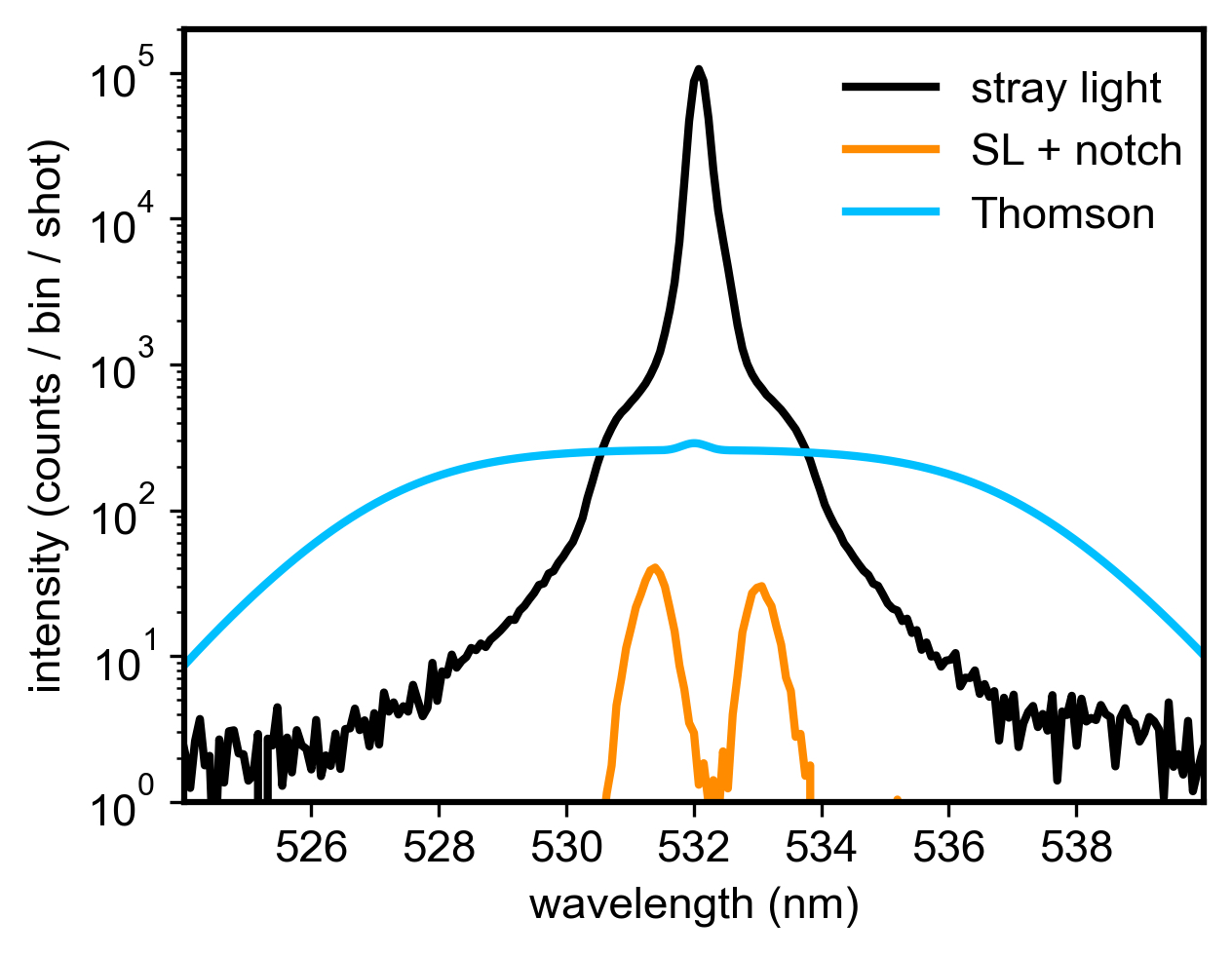}
    \caption{\label{fig:calibration} The stray light spectrum (black) is used to measure the instrument function. The profile shows a Gaussian peak with 0.22~nm FWHM and the Loretzian wings caused by diffraction off the gratings. A simulated and absolutely scaled TS spectrum for 7~eV and 4$\times$10$^{16}$ cm$^{-3}$ is shown for comparison (blue line). At 532~nm the stray light is three orders of magnitude higher in amplitude than the TS signal. The orange curve shows the leakage of the stray-light through the 1.5~nm notch filter. The notch reduces the central stray light ($\lambda_i$) by five orders of magnitude and the wings by three. The notch filter suppresses the stray light to well below 1$\%$ of the TS signal except for wavelengths close to $\lambda_i$.}
\end{figure}

Fig. \ref{fig:calibration} shows the spectrum of stray light, measured without the notch, caused by the probe laser scattering off the target and other surfaces inside the chamber. Since the laser line width is negligible, the profile is a measure of the instrument function. The spectrum is averaged over 300 shots to increase SNR in the wings. The instrument profile is the convolution of the Gaussian contribution due to the aberrated slit width, and the Lorentzian profile caused by diffraction of the three gratings.
The notch reduces the intensity of the central 532~nm wavelength by five orders of magnitude (orange line) and the wings by three orders of magnitude.
The notch reduces the intensity of the wings also outside its 1.5 nm width, since it removes 532 nm light before it diffracts off the final two gratings.
The total extinction ratio of the notch, integrated over all wavelength was $2.0\times 10^4$, equivalent to an optical density of OD~4.3. A simulated and absolutely scaled TS spectrum for $4\times 10^{16}$~cm$^{-3}$ and 7~eV is shown for comparison (blue line). Without the notch, the stray light exceeds the scattering signal by three orders of magnitude, and the stray light amplitude in the wings is no less than 10$\%$ of the TS intensity across all wavelengths. The notch filter suppresses the stray light to well below 1$\%$ of the TS signal except for wavelengths closest to $\lambda_i$.

\section{\label{sec:calibration}Absolute spectrometer calibration}
For laser light scattering off microscopic particles, the total number of counts (photo-electrons) measured by a detector is
\begin{equation}
N = \underbrace{ \overbrace{ \frac{\tau_L \cdot I_L}{h\nu_i}}^{\rm{laser}}  \overbrace{  \Delta V  \Delta \Omega}^{\rm{collection}} \cdot  \overbrace{ \mu\ \eta \ G}^{\rm{optics}}}_{k}  \cdot\ n \cdot \frac{d\sigma}{d\Omega}\ ,
\end{equation}
where $n$ is the density of scattering particles and $d\sigma/d\Omega$ is the differential cross section. 
The experimental throughput parameter $k$  depends on the probe laser, the collection optics and scattering volume, transmission through all optical components, and the camera response.
Here $I_L$ is the probe laser intensity in the scattering volume $\Delta V$, and $h\nu_i$ is the energy of one laser photon. Only a fraction of the photons scattering into the solid angle $\Delta \Omega$ make it to the detector, depending on the optical transmission $\mu$ through the lens, optical fiber, and spectrometer. Lastly, photons are converted to photo-electrons with quantum efficiency $\eta$, which are multiplied by the MCP with gain G before being digitized on the CCD.

The total number of counts $N_T$ in a Thomson scattering spectrum is proportional to the electron density $n_e$
\begin{equation}
    N_T = k\cdot n_e \frac{d\sigma_T}{d\Omega}\ ,
\label{tscount}
\end{equation}
where $d\sigma/d\Omega = r_e^2$ for scattering perpendicular to the probe beam, and $r_e=2.818\times 10^{-15}$ m is the classical electron radius \cite{froula12}.
If $k$ is known, $n_e$ can be deduced from the measured counts. Since $k$ is difficult to calculate accurately \textit{a priori}, it is easier to determine the constant {\it in situ}.
This can be done using the relative signal intensities of Rayleigh \cite{berni1996} or Raman \cite{flora1987} scattering off of a gas. Since stray light and Rayleigh scattered light have the same wavelength ($\lambda_i$), measurements at different pressures and without the notch filter are needed to distinguish between Rayleigh scattered light and stray light.
In this experiment, this calibration is performed using Raman scattering off of gaseous nitrogen at a pressure of 0.86~atm.
Raman scattering has the advantage of producing light outside the notch filter, over a wavelength range and at a signal intensity similar to TS. The disadvantage is that the signal is weak, so spectra must be averaged over thousands of shots to accumulate sufficient photon counts.

\begin{figure}[!ht]
    \centering 
    \includegraphics[width=3.4in]{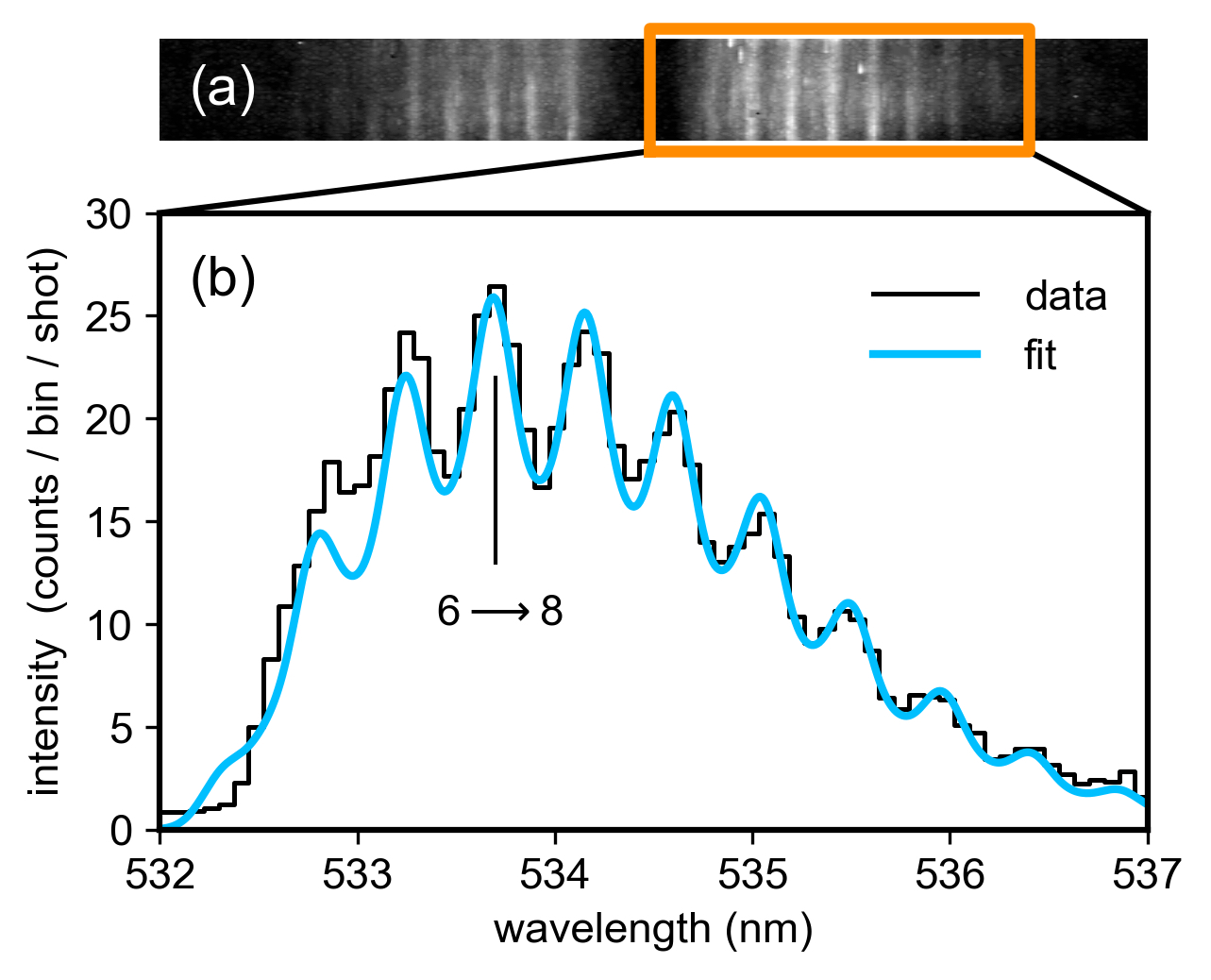}
    \caption{\label{fig:raman} (a) Raw CCD image of a Raman scattering spectrum showing several transition lines around the notch at $\lambda_i$ in the center. 
    (b) Measured spectrum of the red-shifted Stokes lines (black) and comparison with the synthetic fit (blue). Only the bright even J-lines are separated at the spectrometer resolution. The brightest line is the $J=6\rightarrow 8$ transition at 533.7~nm. The area under the fit is used to determine $k$. }
\end{figure}
Fig. \ref{fig:raman}a shows a raw Raman scattering CCD image averaged over 20,000~shots. 
The spectrum consists of a number of spectral peaks on both sides of the 532~nm laser-line, blocked by the notch (in the center of the image). 
Each peak corresponds to a transition from one rotational state to another, induced by the inelastic scattering process, where each state is characterized by the rotational integer quantum number $J$. The energy of a rotational state $J$ is given by $E_J = B\cdot J(J+1)$, where $B=2.48\times 10^{-4}$~eV for nitrogen \cite{penney1974}.
Only transition $J\rightarrow J+2$ (Stokes) and $J\rightarrow J-2$ (anti-Stokes) are allowed. The wavelengths of the red-shifted Stokes lines may be approximated by \cite{vds}
\begin{equation}
    \lambda_{J\rightarrow J+2} \approx \lambda_i + \frac{\lambda_i^2}{hc}B(4J+6)\ .
\end{equation}
The total number of scattered photons in a single Raman line $J$ is given by 
\begin{equation}
    \label{eqn:statecounts}
    N_{J\rightarrow J'} = k\cdot n_J \frac{d\sigma_{J\rightarrow J'}}{d\Omega}\ ,
\end{equation}
where the differential cross section $d\sigma/d\Omega$ varies with state \cite{vds} around a weighted average of $3.8\times 10^{-34}$~m$^2$ .
The density $n_J$ of a rotational state $J$ is
\begin{equation}
    n_J = n_{gas} \frac{g_J(2J+1)}{Q} exp\left(-\frac{E_J}{k_BT}\right)\ ,
\end{equation}
where the partition sum is $Q\approx 9k_B T/B$, and
$g_J$ is a statistical weight factor which is 6 or 3 for even or odd J respectively \cite{herzberg1950}.
The counts in the Raman fine-structure spectrum as a function of wavelength is then
\begin{equation}
    \label{raman}
    I_{\rm{fs}}(\lambda) = k \sum_J  n_J \frac{d\sigma_{J\rightarrow J'}}{d\Omega} \delta(\lambda - \lambda_{J\rightarrow J'})\ ,
\end{equation}
where $\delta(\lambda - \lambda_{J\rightarrow J'})$ is the delta function.
The width of the measured Raman peaks is determined by the instrument broadening, and the synthetic Raman spectrum $I_{\rm{fit}} = I_{\rm{fs}} \circledast I_{\rm{instr}}$ can be  constructed via convolution of the fine structure profile and the experimentally measured instrument function $I_{\rm{instr}}$. 
Fig. \ref{fig:raman} shows the measured Stokes Raman spectrum and comparison with the synthetic fit. The spectral resolution only allows the brighter even J-lines to be separated. The total intensity of all Stokes lines is calculated from the area under the synthetic fit, since transition lines closest to the laser line are partially suppressed by the notch.
Summing equation \ref{eqn:statecounts} over all Stokes J-lines visible at room temperature ($J<25$) gives the total counts in the Stokes spectrum at density $N_{gas}$ from which $k$ can be determined. 
\begin{equation}
N_{Stokes} = k\cdot n_{gas} \cdot 3.82\times 10^{-34} m^2\ .
\end{equation}
We measured a total Stokes signal of $745\pm~50$ counts per shot for a gas density of $n_{gas}=(2.1\pm 0.01) \times 10^{19}$ cm$^{-3}$, corresponding to $k = (9.3\pm 0.7)\times 10^{10}$~m.
The gas density was determined from a pressure measurement at room temperature using a capacitance manometer. 
Misalignment of the collection branch relative to the probe beam decreases the calibration factor $k$.
However, accurate and reproducible alignment can be accomplished by automatically scanning the lens and fiber across the beam as shown in figure 1b to find the peak. Alignment has been found to be stable throughout the day.

\section{\label{sec:results}Results and discussion}
\begin{figure}[!ht]
    \centering 
    \includegraphics[width=3.5in]{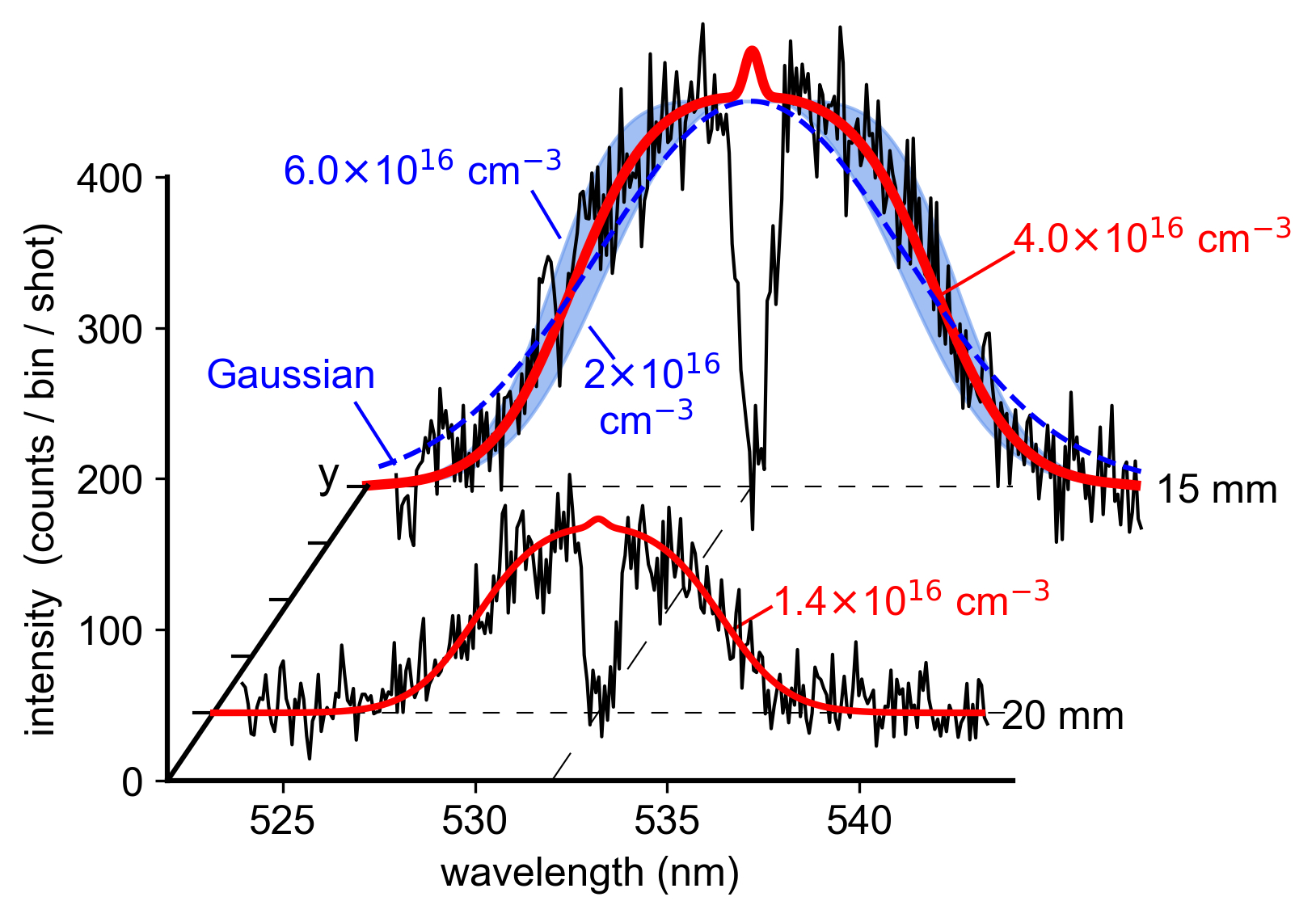}
    \caption{\label{fig:tsspectrum} 
    Measured TS spectra at $y=15$~mm at 150~ns and $y=20$~mm at t=225~ns, with their central bins suppressed by the notch. The 15~mm spectrum is clearly super-Gaussian and can only be fit with a weakly collective spectrum for $n_e=4\times 10^{16}$ cm$^{-3}$ and $T_e=7$~eV. The upper and lower edge of the blue shaded curve show the collective fit for $6\times10^{16}$~cm$^{-3}$ and $2\times10^{16}$~cm$^{-3}$, respectively.
    The 20~mm spectrum is markedly lower in amplitude and exhibits a more Gaussian spectrum consistent with $n_e=1.4\times$ 10$^{16}$~cm$^{-3}$ and $T_e= 4$~eV. 
    }
\end{figure}

Fig. \ref{fig:tsspectrum} compares the TS spectra obtained for a 5~J heater beam at $y=15$~mm and 20~mm from the target. The profiles are recorded at 150~ns and 225~ns respectively.
Each spectrum is averaged over 400 TS shots and an equal number of plasma-only background shots. The central bins of both spectra are suppressed by the notch at $\lambda_i$. Closer to the target, the spectrum exhibits a super-Gaussian profile with a width around 9~nm (FWHM). The spectrum further from the target has a smaller amplitude and area under the curve, a smaller width around 7~nm~(FWHM), and a more Gaussian shape.

%$\alpha = \lambda_i/(4\pi \lambda_D\cdot sin(\Theta/2))$

Thomson scattering can be collective or non-collective, depending on the dimensionless scattering parameter $\alpha = 1 /(k \lambda_D)$, where $k= 4\pi \cdot sin(\theta/2)/\lambda_i$ is the scattering vector and $\lambda_D = \sqrt{\epsilon_0 k_B T_e/n_e e^2}$ is the Debye length. When $\alpha \ll$ 1, the scattering is incoherent (non-collective) and the spectrum reflects the Doppler shifted thermal electron motion, resulting in a Gaussian scattering spectrum. For $\alpha > 1$, the scattering is coherent (collective), and the spectrum is influenced by the interactions between electrons and ions in the plasma. 
The spectrum will be non-Gaussian even if the underlying velocity distribution is Maxwellian as assumed here.
The parameters in this experiment result in scattering parameters near unity and so the spectra range from weakly collective or non-collective with increasing distance from the target~\cite{ross2010, schaeffer2016, seo2018}.
In this experiment the ion feature is blocked by the notch, and the 
TS spectral density function $S(k,\omega)$ that describes the shape of the scattered spectrum is determined only by the electron component. Assuming a quasi-neutral plasma and Maxwellian velocity distribution for the electrons with thermal speed $v_{Te}$, the spectral density function is given by \cite{froula12}
\begin{equation}
    \label{eqn:Skw}
    S(k,\omega) = \frac{2\sqrt{\pi}}{k v_{Te}} \left| 1 - \frac{\chi_e}{\varepsilon} \right| ^2 exp\left( -\frac{\omega^2}{(kv_{Te})^2} \right)\ ,
\end{equation}
where $\varepsilon = 1 + \chi_e + \chi_i$ is the plasma dielectric function, $\chi_e$ and $\chi_i$ are the electron and ion susceptibilities, respectively, $\omega = \omega_s - \omega_i$, $\omega_i$ is the probe beam frequency, and $\omega_s$ is the frequency of the scattered light.
By fitting the experimental spectra with Eq.~\ref{eqn:Skw} convolved with the instrument function, the plasma parameters and $\alpha$ can be inferred. 
The broad shoulders in the 15~mm spectrum in Fig.~\ref{fig:tsspectrum} can only be fitted accurately for temperatures in the range $T_e=(7\pm 0.5)$~eV and densities in the range $(4\pm 2)\times 10^{16}$~cm$^{-3}$, as indicated by the blue shaded area, equivalent to $\alpha=0.61$. 
In this regime the fit to the weakly collective spectrum serves as an independent density diagnostic.  
The total number of counts in the measured TS spectrum (including those suppressed by the notch) is $N_T=(3.3\pm0.2)\times$10$^4$, estimated by calculating the area under the fit which includes the region suppressed by the notch.
Using the experimental throughput parameter $k$ measured from the Raman spectrum, this count is equivalent to a density of $n_e = (4.5\pm 0.6)\times 10^{16}$~cm$^{-3}$, which agrees with the density determined from the collective fit within 10$\%$.
Farther from the target, the fit to the spectrum only yields the temperature, and the density must be determined from the area under the TS spectrum.
The fit to the profile at 20~mm in Fig.~\ref{fig:tsspectrum} corresponds to $1.4\times10^{16}$ cm$^{-3}$, $T_e = 4$~eV, and $\alpha=0.44$.

\begin{figure}[t]
    \centering 
    \includegraphics[width=3.4in]{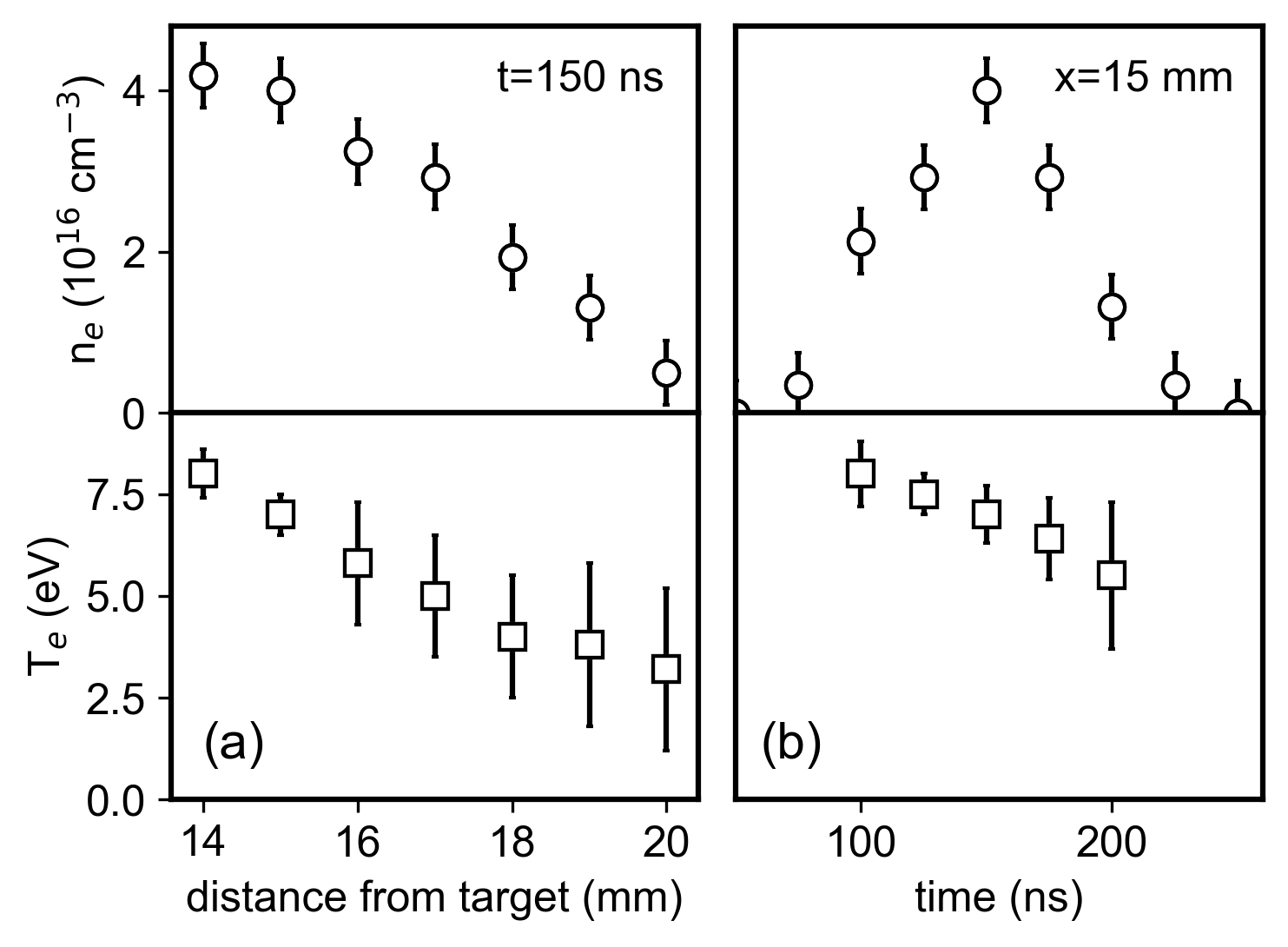}
    \caption{\label{fig:spatialscan} (a) Electron density and temperature as a functions of distance from the target, 150~ns after the heater pulse. 
    (b) Density and temperature as a function of time after the heater pulse at a fixed distance of $y=15$~mm from the target. The data was obtained by scanning the scattering volume automatically along the probe beam. 
    }
\end{figure}

Fig. \ref{fig:spatialscan} shows the plasma parameters as a function of distance from the target at $t=150$~ns (a), and as a function of time at $y=15$~mm (b). This data was recorded automatically by the data-acquisition system in more than $10^4$ laser shots over the course of four hours, by translating the scattering volume or delaying the probe pulse.
Both the density and temperature fall off nearly linearly with distance from the target, consistent with an isentropic fluid model \cite{schaeffer_jap}.
Since the density falls off faster with distance from target than the temperature, the scattering parameter $\alpha \sim \sqrt{n_e/T_e}$ also decreases with distance. 
Data closer to the target than $\sim$12~mm was dominated by intense plasma emission lines (not shown), including neutral carbon lines at 526.9~nm and 528.8~nm, with amplitudes of up to $2\times10^3$ counts and several times higher than the TS signal. The increased shot noise makes it challenging to fit to the TS signal for these spectra.

At a distance of 15~mm from the target the peak plasma density arrives 150~ns after the heater pulse (Fig. \ref{fig:spatialscan}b), consistent with a bulk plasma flow velocity of 100~km/s. Scattered light can be detected as early as 75~ns after the heater beam, so the leading edge of the plasma moves at least at 200~km/s, consistent with previous spectroscopic measurements \cite{dorst2020}.
Scattering spectra in LPPs produced with a higher-energy 10~J heater beam show slightly higher densities and temperatures as expected \cite{schaeffer_jap}. At a distance of 20 mm from the target at 225 ns we measured a density and temperature of $1.7\times 10^{16}$~cm$^{-3}$ and 6~eV, respectively. Closer to the target, at 15~mm, the 10~J spectra at 150~ns are dominated by plasma emission lines (not shown), but the TS fit is consistent with $n_e=(5.5\pm1)\times10^{16}$~cm$^{-3}$ and $T_e=10 \pm$2~eV.

\section{\label{sec:summary}Summary}
We have developed a high-repetition rate Thomson scattering diagnostic that allows auto-alignment and translation of the scattering volume over an arbitrary shaped spatial pattern, as the LPP is produced repeatedly at 1 Hz for thousands of shots.
Using this technique we have measured $n_e$ and $T_e$  in an exploding laser plasma at distances of several centimeters from the target with spatial and temporal resolution. 
Unshifted stray light due to probe beam scattering off optics and surfaces inside the vessel has been effectively mitigated using a triple-grating-spectrometer and notch filter with an extinction ratio of 10$^5$. 
These measurements demonstrate that a collection branch design that sacrifices optical extent at the source in order to reduce beam pointing sensitivity makes it possible to automatically translate the scattering volume using motorized stages without the need for realignment.
Electron densities obtained by fitting the data with the weakly collective spectral density function agree well with densities obtained from an absolute irradiance calibration via Raman scattering. 
These first proof-of-concept measurements were hampered by the very low probe beam energy (50~mJ), which results in a TS signal close to the detection limit for these plasma parameters. 
Near the target or in plasmas produced with a 10 J heater beam, the spectra are dominated by carbon emission lines with intensities five times the TS signal. Scattering spectra are therefore averaged over hundreds of laser shots for each spatial position to improve SNR.
A future iteration of this diagnostic will use a higher energy probe beam ($E_{\rm{pulse}}=0.7$~J) to increase the number of TS photons while keeping the plasma emission background constant. Per Eq.~\ref{snr} this would increase the SNR by more than a factor of 5 for the case $N_T=N_{bg}$. In spectra where the plasma emission now exceeds the TS signal ($N_{bg}= 5\cdot N_T$), the SNR would increase by a factor of 8. TS spectra can then be collected at these densities in as few as 10-20~laser shots.
In combination with an additional stepper-controlled stage to translate the final probe-beam turning mirror along the $\hat{x}$ axis, this technique will 
allow point TS measurements in laser plasmas on a two-dimensional grid for the first time, and produce 2D density and temperature maps. 
While the current low probe beam energy requires too many shots to feasibly perform 2D scans, this limitation will be rectified with a higher energy laser. 

\section{Data availability statement}
The data that support the findings of this study are available from the corresponding author upon reasonable request.

\begin{acknowledgments}
This work was supported by the Department of Energy under contract numbers DE-SC0019011 and DE-SC0021133, by the Defense Threat Reduction Agency and Lawrence Livermore National Security LLC under contract number B643014, by the National Nuclear Security Administration Center for Matter Under Extreme Conditions under award number DE-NA0003842, and by the National Science Foundation Graduate Fellowship Research Program under award number DGE-1650604. We thank NIWC Pacific and Curtiss-Wright MIC for help with the slab laser system.
\end{acknowledgments}

\bibliographystyle{unsrtnat}
\bibliography{bibliography.bib}

\end{document}